\documentclass[9pt,twocolumn,twoside, superscriptaddress]{revtex4-2}

\usepackage[markup=nocolor, authormarkupposition=left]{changes}
\usepackage{soul}
\usepackage[utf8]{inputenc}
\usepackage[english]{babel}
\usepackage{amsmath,amsfonts,amssymb}
\usepackage[T1]{fontenc}
\usepackage{url}
\usepackage{changes}
\usepackage{amsmath}
\usepackage{amsfonts}
\usepackage{amssymb}
\usepackage{siunitx}
\usepackage{epstopdf}
\usepackage{graphicx}
\graphicspath{{./Figures/}}
\usepackage[
bookmarks=true,
colorlinks=true,
linkcolor=blue,
urlcolor=blue,
citecolor=blue,
pdftex,
bookmarks=true,
linktocpage=true, 
hyperindex=true
]{hyperref}

\graphicspath{{./Figures/}}
\usepackage{changes}
\usepackage{soul}
\newcommand{\SiN}[0]{Si$_3$N$_4$~}
\newcommand{\SiO}[0]{SiO$_2$}

\begin{document}

\title{Frequency agile photonic integrated external cavity laser}
	
\author{Grigory Lihachev}
\email{These authors contributed equally}
\affiliation{Institute of Physics, Swiss Federal Institute of Technology Lausanne (EPFL), CH-1015 Lausanne, Switzerland}
\affiliation{Center of Quantum Science and Engineering (EPFL), CH-1015 Lausanne, Switzerland}

\author{Andrea Bancora}
\email{These authors contributed equally}
\affiliation{Institute of Physics, Swiss Federal Institute of Technology Lausanne (EPFL), CH-1015 Lausanne, Switzerland}
\affiliation{Center of Quantum Science and Engineering (EPFL), CH-1015 Lausanne, Switzerland}
\affiliation{Deeplight SA, CH-1015 Lausanne, Switzerland}

\author{Viacheslav Snigirev}
\affiliation{Institute of Physics, Swiss Federal Institute of Technology Lausanne (EPFL), CH-1015 Lausanne, Switzerland}
\affiliation{Center of Quantum Science and Engineering (EPFL), CH-1015 Lausanne, Switzerland}

\author{Hao Tian}
\affiliation{OxideMEMS Lab, Purdue University, 47907 West Lafayette, IN, USA}

\author{Johann Riemensberger}
\affiliation{Institute of Physics, Swiss Federal Institute of Technology Lausanne (EPFL), CH-1015 Lausanne, Switzerland}
\affiliation{Center of Quantum Science and Engineering (EPFL), CH-1015 Lausanne, Switzerland}

\author{Vladimir Shadymov}
\affiliation{Institute of Physics, Swiss Federal Institute of Technology Lausanne (EPFL), CH-1015 Lausanne, Switzerland}
\affiliation{Center of Quantum Science and Engineering (EPFL), CH-1015 Lausanne, Switzerland}

\author{Anat Siddharth}
\affiliation{Institute of Physics, Swiss Federal Institute of Technology Lausanne (EPFL), CH-1015 Lausanne, Switzerland}
\affiliation{Center of Quantum Science and Engineering (EPFL), CH-1015 Lausanne, Switzerland}

\author{Alaina Attanasio}
\affiliation{OxideMEMS Lab, Purdue University, 47907 West Lafayette, IN, USA}

\author{Rui Ning Wang}
\affiliation{Institute of Physics, Swiss Federal Institute of Technology Lausanne (EPFL), CH-1015 Lausanne, Switzerland}
\affiliation{Center of Quantum Science and Engineering (EPFL), CH-1015 Lausanne, Switzerland}

\author{Diego A. Visani}
\affiliation{Institute of Physics, Swiss Federal Institute of Technology Lausanne (EPFL), CH-1015 Lausanne, Switzerland}
\affiliation{Center of Quantum Science and Engineering (EPFL), CH-1015 Lausanne, Switzerland}

\author{Andrey Voloshin}
\email{Corresponding author: andrey.voloshin@epfl.ch}
\affiliation{Institute of Physics, Swiss Federal Institute of Technology Lausanne (EPFL), CH-1015 Lausanne, Switzerland}
\affiliation{Center of Quantum Science and Engineering (EPFL), CH-1015 Lausanne, Switzerland}
\affiliation{Deeplight SA, CH-1015 Lausanne, Switzerland}

\author{Sunil Bhave}
\email{Corresponding author: bhave@purdue.edu}
\affiliation{OxideMEMS Lab, Purdue University, 47907 West Lafayette, IN, USA}

\author{Tobias J. Kippenberg}
\email{Corresponding author: tobias.kippenberg@epfl.ch}
\affiliation{Institute of Physics, Swiss Federal Institute of Technology Lausanne (EPFL), CH-1015 Lausanne, Switzerland}
\affiliation{Center of Quantum Science and Engineering (EPFL), CH-1015 Lausanne, Switzerland}

\maketitle
	
\noindent \textbf{Recent advances in the development of ultra-low loss silicon nitride integrated photonic circuits have heralded a new generation of integrated lasers capable of reaching fiber laser coherence.  
However, these devices presently are based on self-injection locking of distributed feedback (DFB) laser diodes, increasing both the cost and requiring tuning of laser setpoints for their operation. 
In contrast, turn-key legacy laser systems use reflective semiconductor optical amplifiers (RSOA). While this scheme has been utilized for integrated photonics-based lasers, so far, no cost-effective RSOA-based integrated lasers exist that are low noise and simultaneously feature fast, mode-hop-free and linear frequency tuning as required for frequency modulated continuous wave (FMCW) LiDAR or for laser locking in frequency metrology.
Here we overcome this challenge and demonstrate a RSOA-based, frequency agile integrated laser, that can be tuned with high speed, with high linearity at low power.
This is achieved using monolithic integration of piezoelectrical actuators on ultra-low loss silicon nitride photonic integrated circuits in a Vernier filter-based laser scheme.
The laser operates at 1550 nm, features 6 mW output power, 400 Hz intrinsic laser linewidth, and allows ultrafast wavelength switching within 7 ns rise time and 75 nW power consumption.
In addition, we demonstrate the suitability for FMCW LiDAR by showing laser frequency tuning over 1.5 GHz at 100 kHz triangular chirp rate with nonlinearity of 0.25\% after linearization, and use the source for measuring a target scene 10 m away with a 8.5 cm distance resolution. }

\begin{figure*}[t]
	\centering
	\includegraphics[scale=1]{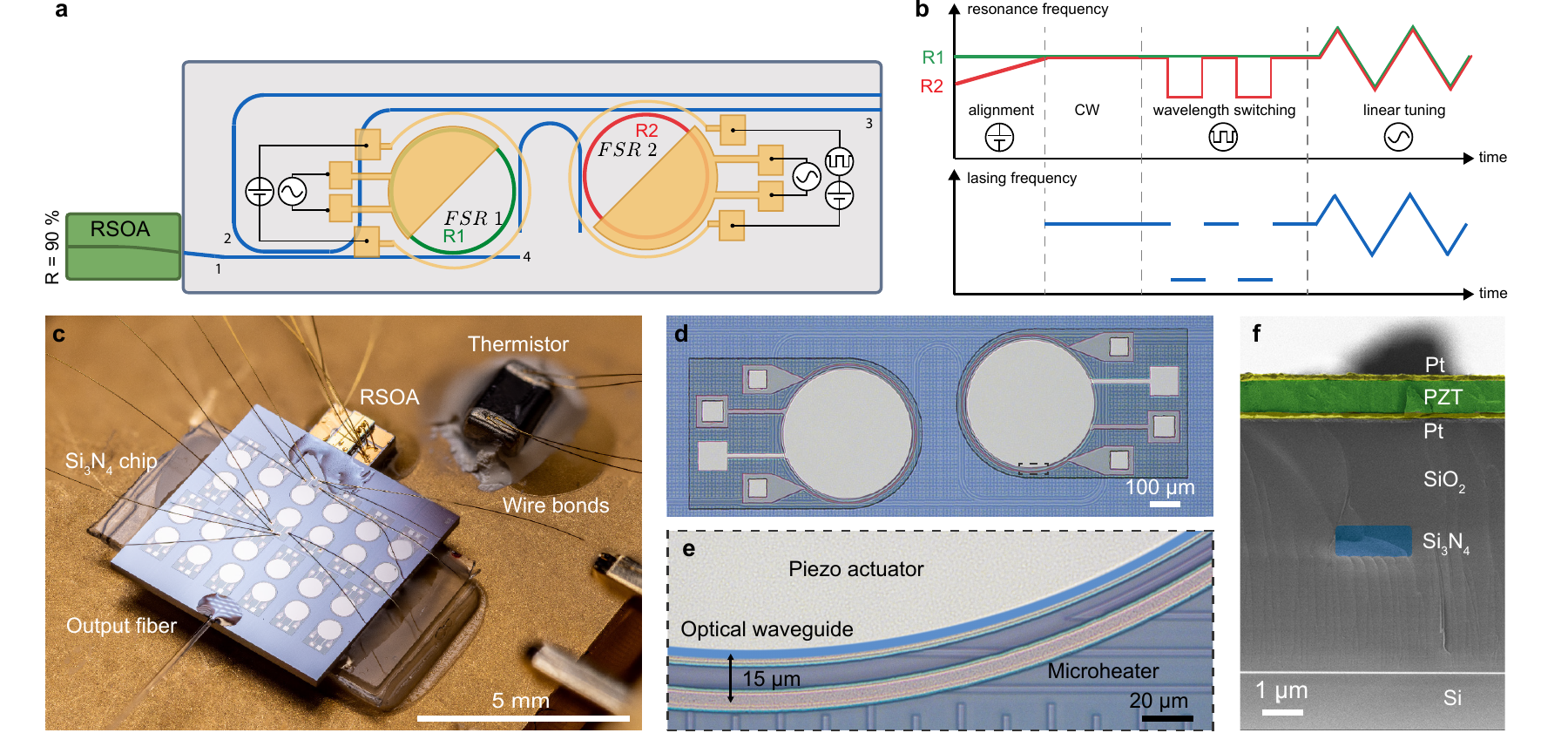}
	\caption{
		\footnotesize \linespread{1}
		\textbf{Integrated external cavity laser.}
		(a)~Schematics of the laser. Reflective semiconductor optical amplifier (RSOA) is edge coupled to \SiN PIC with 2 microrings in Vernier filter configuration.
		The laser frequency is controlled by integrated microheaters and PZT actuators. DC source is used for microheaters, AFG - arbitrary frequency generator.
		(b)~Laser frequency tuning schematic.
		Coarse alignment of resonances of 2 microrings to observe lasing is done with integrated microheaters.
		Triangular voltage signal applied simultaneously to both piezoactuators results in linear laser frequency sweep.
		Applying square signal to a single piezoactuator allows to perform fast wavelength switching. 
		(c)~Photo of a packaged laser in a butterfly package with output fiber, temperature control and wirebonding of all actuators and heaters.
		(d)~Photo of \SiN microrings with integrated PZT actuators and electrodes (yellow). 
		(e)~Zoom in photo showing microheater strip 15 $\mu$m away from \SiN waveguide.
		(f)~Colored SEM image of the sample cross-section, showing the piezoelectric actuator integrated on the \SiN chip.
		The piezoelectric actuator is composed of Pt (yellow), PZT (green) layers on top of \SiN (blue) buried in \SiO~cladding (grey).		
	}
	\label{Fig:Fig1}
\end{figure*}
\section{Introduction}
Narrow linewidth lasers have many applications in metrology, optical sensing \cite{rogers1999distributed}, microwave photonics \cite{marpaung2019integrated,Pillet:08}, optical trapping \cite{Niffenegger2020} and interconnects in datacenters \cite{9057434}.
The frequency agility of a laser, i.e., the ability to tune and precisely control the laser frequency, allows to use them for tight laser locking in quantum applications, fast wavelength switching applications in telecommunications, and in frequency modulated continuous wave (FMCW) LiDAR \cite{dale2014ultra,hecht2018lidar,rogers2021universal,martin2018photonic,isaac2019indium}.
Over the past two decades, major advances have been made in silicon-based integrated lasers, where both hybrid and heterogeneous integration of III-V have enabled compact lasers that are already commercially used in datacenter transceivers \cite{Komljenovic:16,Tran:2020}.  
Yet, these laser sources have not obtained a phase noise performance that is on par with legacy laser systems, notably continuous wave fiber lasers.
Recent advances in low-loss silicon nitride integrated photonics have heralded a new class of integrated lasers that can outperform legacy lasers in terms of phase noise. 
Using distributed feedback (DFB) diode self-injection locking to integrated microresonators with low confinement \SiN waveguides, a laser surpassing the coherence of a fiber laser has been demonstrated \cite{li2021reaching,Jin2021}, while tightly confining waveguides have enabled lasers that achieve fiber laser phase noise with unprecedentedly fast frequency actuation that is flat over MHz bandwidth \cite{lihachev2022low}.
These advances have been made possible by ultra-low loss silicon nitride integrated photonic circuits.
Over the past decade, silicon nitride tightly confining waveguides (with heights > 700 nm) have obtained propagation loss below 3 dB/m and have become a photonic integrated circuit (PIC) platform that is commercially available via foundry, in tandem with weak confinement platforms \cite{Jin2021}. 
Low propagation loss and high fabrication yield \cite{liu2021high} have enabled novel functionality ranging from soliton microcombs \cite{doi:10.1126/science.abh2076}, traveling wave parametric amplifiers \cite{Riemensberger2022} to Erbium doped amplifiers \cite{doi:10.1126/science.abo2631}. 
Moreover, silicon nitride PICs have been integrated monolithically with piezoelectrical actuators, allowing flat, fast (MHz bandwidth) low power (nano-Watt) and linear tuning, resulting in frequency agile low noise lasers \cite{lihachev2022low} and fast tunable soliton frequency combs \cite{liu2020monolithic}.
While these recent low noise integrated laser demonstrations have allowed integrated photonic lasers to reach the 'gold standard' of fiber laser coherence, these schemes require self-injection locking, which requires careful setpoint and relay on DFB diode which requires grating manufacturing on the III-V die.
In contrast, many commercial lasers utilize external cavities based on reflective semiconductor optical amplifiers (RSOA) or gain chips. 
External cavity lasers (ECL) with feedback circuits implemented in photonic integrated circuits demonstrated significant progress in recent years \cite{Belt:14,Huang:19,duan2014hybrid,tran2019tutorial}.
PICs with double-ring Vernier filters provide frequency-selective reflection.
Lasers with such Vernier filters have significantly matured with the realization of filters in silicon \cite{Tran:2020,doi:10.1063/1.4915306}, silicon nitride \cite{photonics7010004} or other material platforms \cite{Radosavljevic:18,zhang2021integrated} and reached sub-kHz intrinsic laser linewidths \cite{fan2020hybrid,Tran:2020}.
High-frequency modulation speed (exahertz/s) and switching speed of up to 50 MHz have only been demonstrated recently in Vernier-based integrated lasers based on lithium niobate \cite{Li2022}, which are however compounded by phase noise that is above that achieved with silicon nitride \cite{fan2020hybrid}.
Here, we report a hybrid integrated ECL based on RSOA and Vernier filters in low loss \SiN PIC enhanced with integrated piezoelectric actuators, which constitutes a low cost solution, alleviating the use of DFB, while enabling high coherence and fast (MHz bandwidth), linear and low power frequency tuning.
Several application areas would benefit from such low noise, frequency agile ECL based on low-cost RSOA. Examples are fast wavelength switching for data centers \cite{guan2018widely}, which has been studied in digital supermode distributed Bragg reflector (DS-DBR) \cite{Yoo:17}, Vernier-tuned distributed Bragg reflector (VT-DBR) \cite{9563804} and DFB \cite{ueno2015fast} configurations with mode-hop-free wavelength tuning \cite{vanRees:20} and wavelength switching \cite{srinivasan2015coupled,li2018tunable} demonstrated using integrated heaters with speeds up to 10 kHz.
Another application is FMCW LiDAR, where high laser frequency tuning linearity and low frequency noise of the laser are crucial to measuring distance and velocity at mid to long ranges \cite{Harris:98,Behroozpour:17}.
Despite significant progress, the fast and linearly tunable integrated laser requires either a DFB laser \cite{dilazaro2018large} with e-beam lithography fabrication steps or MEMS-VCSELs with additional linearization \cite{Okano:20,Feneyrou:17}. 

\section{Photonic integrated chip design and characterization}
We implement a laser with an external cavity realized using tightly confining \SiN photonic integrated circuits with double-ring Vernier filters \cite{tran2019tutorial}.
Fig. \ref{Fig:Fig1}(a) shows the design of the photonic integrated circuit and the schematic of the laser. 
A reflective semiconductor optical amplifier (RSOA, Thorlabs SAF1126C) with 90\% back facet reflection is edge coupled to \SiN PIC with a directional coupler (splitter) and two microrings.
The optical path is marked with blue arrows.
The laser output can be measured at ports 2, 3, 4 of the splitter.
The light after the splitter is coupled to microrings with diameters of 472 $\mu$m and 465 $\mu$m via symmetrical input and drop-ports with 350 nm bus waveguide - ring waveguide gaps. 
Fig.~\ref{Fig:Fig1}(f) shows a cross-section of the chip with the following material stack: Si carrier (black), \SiN waveguide with 2200 nm width and 900 nm height, top SiO$_2$ cladding with 2.5 $\mu$m height (grey), bottom and top Pt electrodes (yellow), PZT 1000 nm (green). 
\SiN waveguides are fabricated using Damascene reflow process \cite{Pfeiffer:18}. 

We developed the actuation of both rings of a Vernier structure utilizing the stress-optical effect and the thermo-optic coefficient.
A monolithically integrated actuator (see Fig. \ref{Fig:Fig1}(d)) comprises a piezo-electrical PZT actuator \cite{lihachev2022low,Wang:22} for fast actuation and a microheater for alignment of a Vernier filter and wide wavelength tunability.
Microheaters are fabricated in the bottom (ground) electrode layer which can be accessed through an opening of the PZT layer for wirebonding.
Microheaters are placed 15 $\mu$m away from the ring \SiN waveguide as depicted in Fig. \ref{Fig:Fig1}(e).
Such heater position is dictated by the fabrication constraint on the minimum distance between the bottom electrode for the PZT actuator and the heater strip as well as on the optimization of the PZT actuator position to increase the stress-optic tuning efficiency (166 MHz/V for our device).
The radius of the top electrode for the PZT actuator is equal to that of the \SiN ring.

\begin{figure*}[t]
	\centering
	\includegraphics[scale=1]{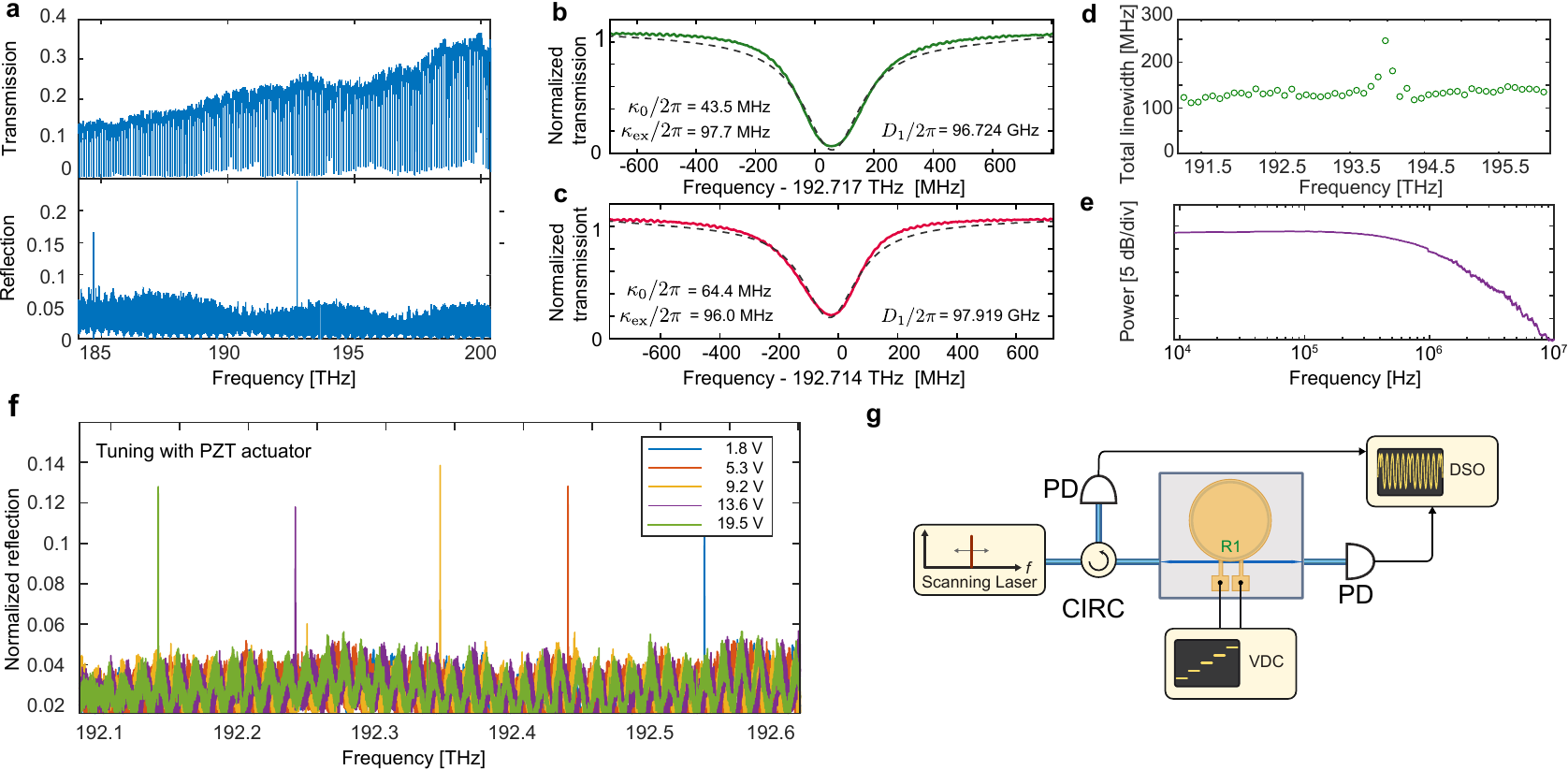}
	\caption{
		\footnotesize \linespread{1}
		\textbf{Integrated external cavity laser.}
		(a)~Frequency dependent cavity transmission and reflection (d) measured with frequency comb assisted calibration spectroscopy. Clear reflection peaks are visible when resonances of 2 rings are aligned.
		(b)~Cavity transmission (blue) and lorentzian fit (red) of the resonance of 1st microring with FSR of 96.7 GHz and intrinsic linewidth of 43.5 MHz with symmetric input/drop-port coupling. 
		(c)~Cavity transmission (blue) and lorentzian fit (red) of the resonance of 2nd microring with FSR of 97.9 GHz and intrinsic linewidth of 96 MHz.  
		(d)~Distribution of total linewidth of R1 resonances in the RSOA amplification band (1528-1568 nm).
		(e)~Optomechanical S$_{21}$ response of the packaged device with Vernier filter showing flat actuation bandwidth up to 960 kHz.
		(f)~Normalized reflection of the Vernier filter at the different voltages applied to the integrated PZT actuator. Increasing voltage by 8 V shifts the cavity resonance by 1.2 GHz and allows to shift the reflection peak by 1 FSR (97 GHz).
		(g) Schematic of the experiment to measure transmission and reflection of Vernier filter at different voltages applied to PZT actuator. PD - photodiode, CIRC - optical circulator, DSO - digital storage oscilloscope.
	}
	\label{Fig:Fig2}
\end{figure*}

First, we perform passive characterization of a PIC-based \SiN Vernier filter using the frequency comb-assisted calibration spectroscopy \cite{del2009frequency}.
Frequency dependent cavity transmission and reflection presented in Fig. \ref{Fig:Fig2}(a) show 10\%-35\% transmission measured for ring R1 in port 4 of the splitter.
Reflection peaks up to 25\% in power occur when resonances of 2 microrings are aligned by the microheater.
Fig. \ref{Fig:Fig2} (b,c) show the cavity transmission (green, red) and Lorentzian fit (dashed) of the resonances of microrings R1 and R2 with FSR$_1=96.7$ GHz, FSR$_2=97.9$ GHz, and intrinsic linewidths $\kappa_0/2\pi$ of 43.5 MHz and 64.4 MHz with symmetric input/drop-port coupling of $\kappa_{ex}/2\pi = 96.0$ MHz correspondingly.
Fig. \ref{Fig:Fig2}(d) shows the total cavity linewidth for resonances of microring R1 in the RSOA amplification band, that can be used for lasing.
The Vernier FSR \cite{tran2019tutorial} calculated as $\text{FSR}_1 \cdot \text{FSR}_2/(\text{FSR}_1-\text{FSR}_2)$ is 8.7 THz for our device \cite{tran2019tutorial}, which leads to a simple alignment for a single mode emission regime. Then we characterized the reflection of a PIC-based Vernier filter actuated by PZT actuators. Before driving piezoactuators, we pole the ferroelectric PZT material to align domains and increase tuning efficiency by applying 25 V DC to the actuator for several seconds \cite{4805529}.
To maintain the direction of the polarization we apply only positive voltages in experiments.
We apply different voltages to a single PZT actuator keeping the temperature of RSOA and microheaters fixed. 
In fig. \ref{Fig:Fig2}(f) we show the normalized reflection of the Vernier filter at the different voltages applied to the integrated PZT actuator. 
Increasing voltage by 8 V tunes the cavity resonance by 1.2 GHz, equal to the FSR mismatch of two rings, and shifts the reflection peak by 1 FSR (97 GHz).
The saturation voltage of the PZT actuator is around 35 V for our device, which allows us to switch the Vernier filter frequency up to 4 FSRs (400 GHz) using only the PZT actuator.
In all measurements the current draw on the actuator was less than 20 nA, resulting in low power consumption of 75 nW for the wavelength switching.

\section{Photonic integrated laser characterization}

Fig. \ref{Fig:Fig1}(b) presents a laser frequency tuning schematic based on PICs described above.
First, we align a pair of resonances of two rings to observe lasing by applying DC voltage to one of microheaters.
Then, a sawtooth voltage signal applied simultaneously to both piezoactuators results in a linear laser frequency sweep.
Applying a square signal to a single piezoactuator allows the performing of fast wavelength switching. In our laser design, we do not have an on-chip phase shifter, instead, we use phase shifting capabilities of the active section by changing the RSOA current.
With a fixed RSOA injection current, the maximum tuning range of the laser in a single-mode regime is limited by 3 GHz.
We perform hybrid packaging of the laser by mounting and gluing the RSOA and \SiN chip with output fiber, placing a temperature control and wirebonding of all actuators and heaters in a custom butterfly package as depicted in Fig. \ref{Fig:Fig1}(c). Hybrid packaging improves the long-term laser stability and reduce laser frequency noise below 1 kHz offsets by eliminating acoustic instabilities inherent to the unpackaged optical setup.

\begin{figure*}[t]
	\centering
	\includegraphics[scale=1]{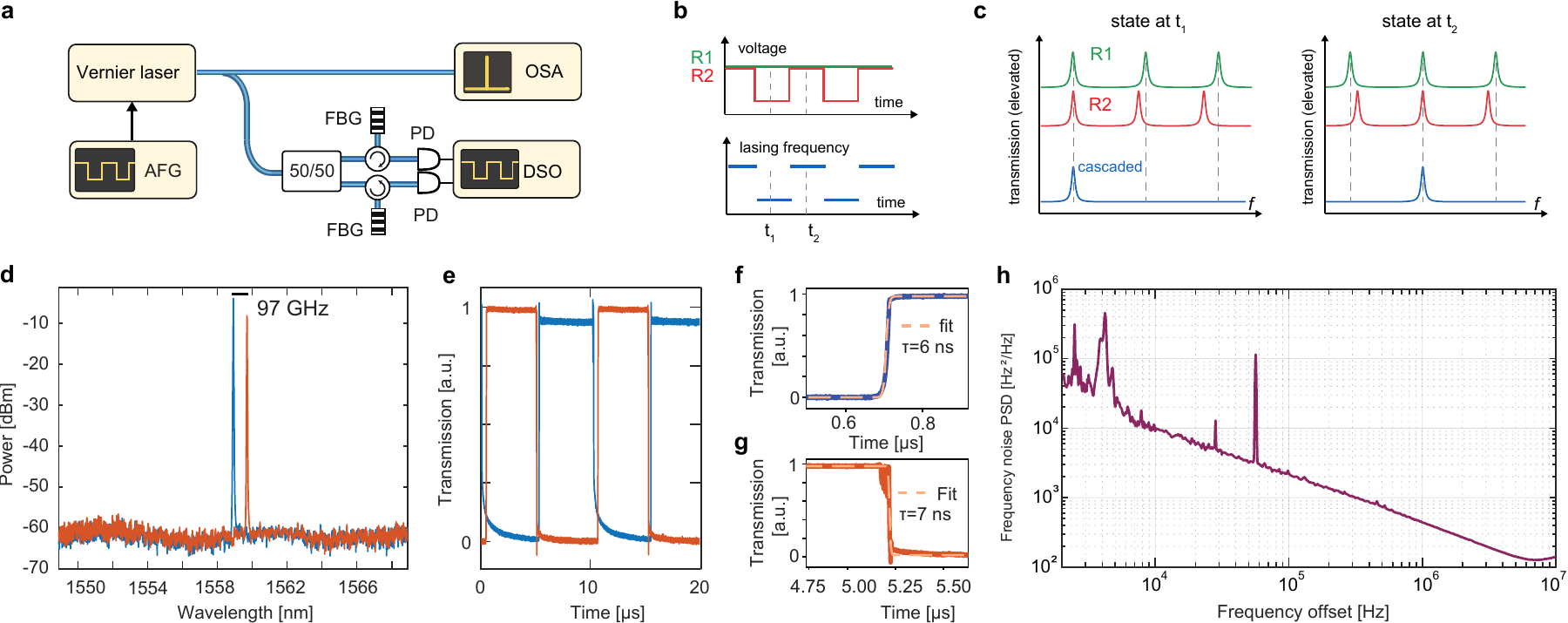}
	\caption{
		\footnotesize \linespread{1}
		\textbf{Fast wavelength switching.}
		(a)~Experimental setup. DUT - laser device under test, AFG - arbitrary function generator, OSA - optical spectrum analyzer, FBG -Fiber Bragg grating, PD - photodetector, DSO - digital oscilloscope.
		(b,c)~The schematic of the fast wavelength switching using single PZT actuator and square driving voltage.
		(d)~Optical spectra showing laser wavelength switching of 97 GHz between two states: red and blue.
		(e)~Measured transmission in both channels (red and blue correspond to (d)), demonstrating laser wavelength switching at 100 kHz rate.
		(f,g)~Transmission curve (red) and its fit (blue) reveals the rise time of 7 ns and the fall time of 6 ns.
		(h)~Single-sided PSD of frequency noise of the hybrid integrated external cavity laser. 
	}
	\label{Fig:Fig3}
\end{figure*}

We analyze the frequency noise of the hybrid laser. Towards this end, we performed a heterodyne beatnote spectroscopy \cite{duthel2009laser} beating a reference external cavity diode laser (free-running Toptica CTL) with our ECL. 
The beatnote of the two laser frequencies was detected on a fast photodiode (Finisar XPDV2120RA), and its electrical output was then sent to an electrical spectrum analyzer (Rohde \& Schwarz FSW43).
The recorded data for the in-phase and quadrature components of the beatnote were processed by Welch's method \cite{welch1967use} to retrieve the single-sided phase noise power spectral density $S_{\phi \phi}$ that was converted to frequency noise $S_{ff}$. 
The frequency noise of the reference laser was determined by a separate beatnote measurement with a commercial ultrastable laser (Menlo ORS). 
Fig.~\ref{Fig:Fig3}(h) shows the single-sided power spectral density (PSD) of the laser frequency noise.
The frequency noise of the Vernier laser is limited by the reference laser below 2 kHz, then it follows $1/f$ slope at offsets 10 kHz - 100 kHz and $1/f^{1/2}$ slope at offsets 100 kHz - 4 MHz.
The laser frequency noise reaches a value (white noise floor) of 127 Hz$^2$/Hz at a 6 MHz offset, after multiplying by 2$\pi$ corresponding to the intrinsic laser linewidth of 400 Hz.
The hybrid integrated ECL exhibits 6 mW of output power in fiber at 1567 nm and the side mode suppression ratio (SMSR) of 50 dB. 

\begin{figure*}[t]
	\centering
	\includegraphics[scale=1]{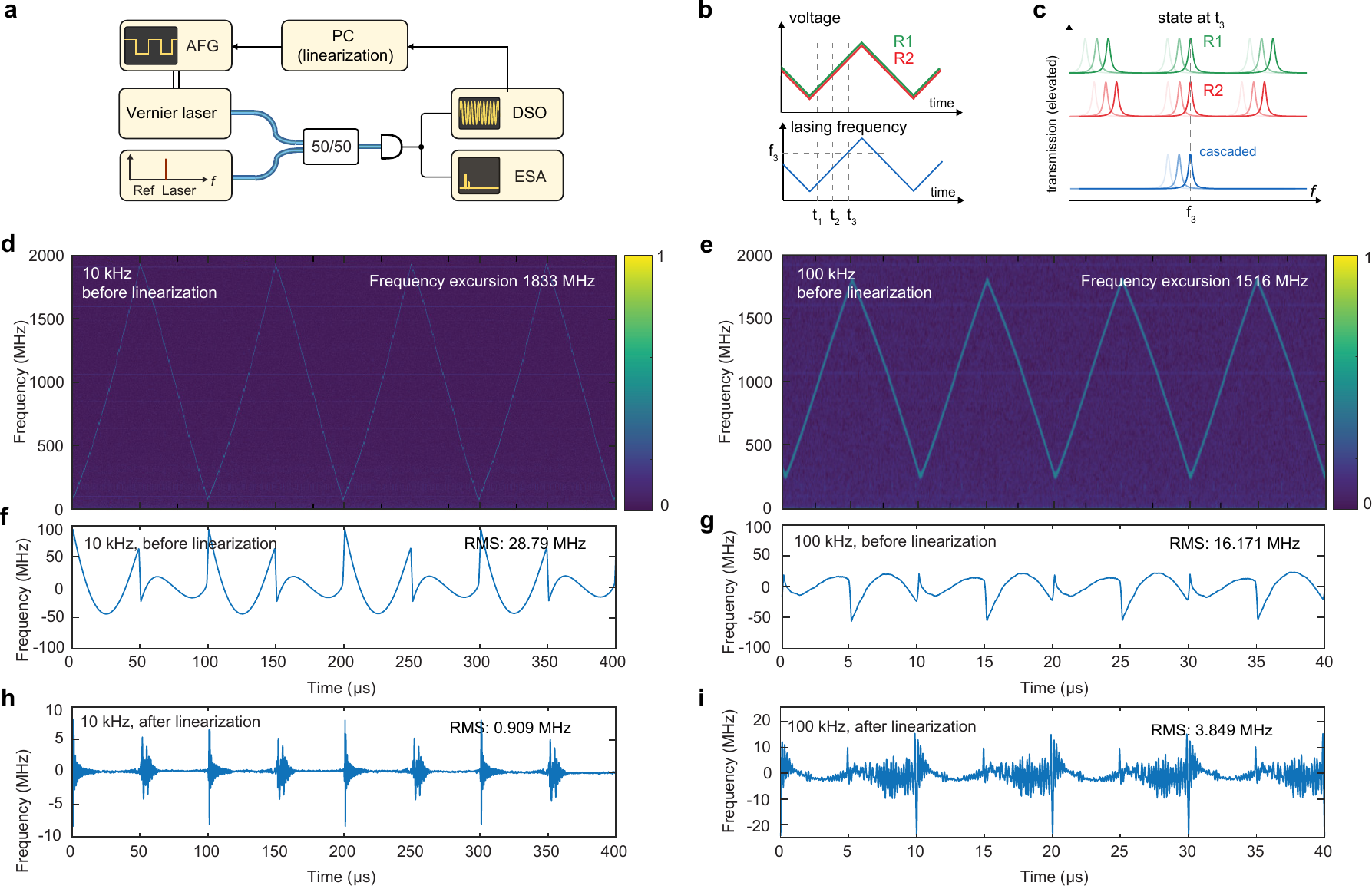}
	\caption{
		\footnotesize \linespread{1}
		\textbf{Fast linear frequency tuning.}
		(a)~Experimental setup. DUT - laser device under test, AFG - arbitrary function generator, OSA - optical spectrum analyzer, PD - photodetector, DSO - digital storage oscilloscope.
		(b,c)~The schematic of the linear tuning by applying the same driving triangular signal to both PZT actuators.
		(d,e)~Laser frequency tuning spectrograms with the corresponding nonlinearities (f,g) at 10 kHz and 100 kHz chirp frequencies upon applying 11 Vpp non-corrected ramp signal from AFG to PZT actuators.
		(h,i)~Tuning nonlinearities after 10 iterations of predistortion compensation. At 10 kHz rate relative RMS is 0.05\%.  
	}
	\label{Fig:Fig4}
\end{figure*}

\section{Fast wavelength switching}
Next we demonstrate fast wavelength switching of the ECL using only PZT actuators.
A conceptual schematic of the experiment is presented in Fig.~\ref{Fig:Fig3}(b,c) and the experimental setup in Fig.~\ref{Fig:Fig3}(a).
We split the output of the laser in two paths with a fiber splitter, and install in each arm optical circulators followed by fiber Bragg gratings (FBG) notch filter with 8 GHz bandwidth.
FBGs are centered at wavelengths separated by 97 GHz, which allows measuring both channels using photodetectors (NewFocus Model 1811, 125 MHz bandwidth) in port 3 of circulators.
To switch the wavelength of the laser we align the Vernier filter with a microheater and then actuate only one ring with the PZT actuator, applying a square signal from 0V to 8V at a 100 kHz rate.
The driving square signal from the AFG and its fitting are presented in the SI.
Fig.~\ref{Fig:Fig3}(e) depicts the transmission power in both channels, showing wavelength switching of 97 GHz at 100 kHz rate.
Optical spectra of the laser are presented in Fig.~\ref{Fig:Fig3}(d), showing two emission wavelengths separated by an FSR.
To determine switching speed we fit the rise and fall transmission curves with the hyperbolic tangent function $\sim \tanh\left((t-t_0)/\tau\right)$, where $\tau$ stands for the switching time, and $t_0$ for the step time offset. Fitting yields 10–90\% rise time of 7 ns and fall time of 6 ns (see Fig. \ref{Fig:Fig3}(f,g)).
Such fast wavelength switching and low measured actuator power consumption per switching of only 75 nW demonstrate an advantage of ECL with integrated PZT actuators over microheaters with typical switching times of hundreds of nanoseconds.
The demonstrated values of switching time are on par with the recently demonstrated Vernier laser based on lithium niobate slab waveguides and Pockels effect-based tuning \cite{Li2022}.

\section{Fast frequency tuning and linearization}
Finally, we characterize the frequency agility of our hybrid laser.
The optomechanical S$_{21}$ response of the packaged \SiN device is presented in Fig.~\ref{Fig:Fig2}(e).
The electro-optical response of the PZT actuator is flat up to 960~kHz, indicating the wide span of frequencies where the Vernier laser wavelength can be efficiently modulated.
The concept of the fast frequency tuning experiment is presented in Fig.~\ref{Fig:Fig4}(b,c).
The Vernier filter is aligned with microheaters to achieve single-mode lasing at 1559 nm.
We apply a triangular signal with 11 V$_\mathrm{p-p}$ amplitude from an arbitrary function generator (AFG) Tektronix AFG3102 simultaneously to both PZT actuators at chirp frequencies 10~kHz and 100~kHz (see the experimental setup in Fig. \ref{Fig:Fig4}(a)).
The chirped laser output frequency is measured by a heterodyne beatnote with a reference ECDL on a fast photodetector. 
We define the chirp nonlinearity as the root mean square (RMS) deviation of the measured frequency tuning curve from a perfect symmetric triangular ramp signal that is determined with least-squares fitting.
Fig.~\ref{Fig:Fig4}(d,e) presents the processed laser frequency spectrograms with the corresponding nonlinearities (Fig.~\ref{Fig:Fig4}(f,g)) at 10~kHz and 100~kHz ramping frequencies, respectively.
Frequency excursions of 1833~MHz, corresponding to 166~MHz/V tuning efficiency, with RMS relative nonlinearities of 1.5\% at 10~kHz ramping frequency and frequency excursion of 1516~MHz with 1\% nonlinearity at 100~kHz are demonstrated using PZT actuator without any additional linearization.
The key requirements for various photonic sensing applications, e.g., FMCW LiDAR at medium to long ranges, are conflicting: very high tuning linearity and a low laser frequency jitter \cite{Behroozpour2017}.
To improve the RMS tuning nonlinearity of our ECL, we combine two algorithms for tuning voltage correction: frequency response inversion and iterative signal correction.
First, the frequency response of the actuator is determined by applying a short Gaussian voltage pulse is sent to both actuators, and the laser frequency response is measured using heterodyne beat with the reference laser on a fast PD and oscilloscope with 2 GHz acquisition bandwidth. 
Both the initial voltage ramp and the correction in each iteration of the algorithm are multiplied by the inverse frequency response \cite{Feneyrou:17}. 
After 10~iterations, at 10~kHz modulation frequency, the achieved RMS nonlinearity is as low as 0.9~MHz (relative nonlinearity 0.05\%), which degrades for 100~kHz tuning rate to 3.85~MHz (0.25\%) (see Fig.~\ref{Fig:Fig4}(h,i)), thus, demonstrating an improvement factor of 30.

\section{Optical coherent ranging using the hybrid integrated laser}
\begin{figure*}[t]
	\centering
	\footnotesize \linespread{1}
	\includegraphics[width=0.7\textwidth]{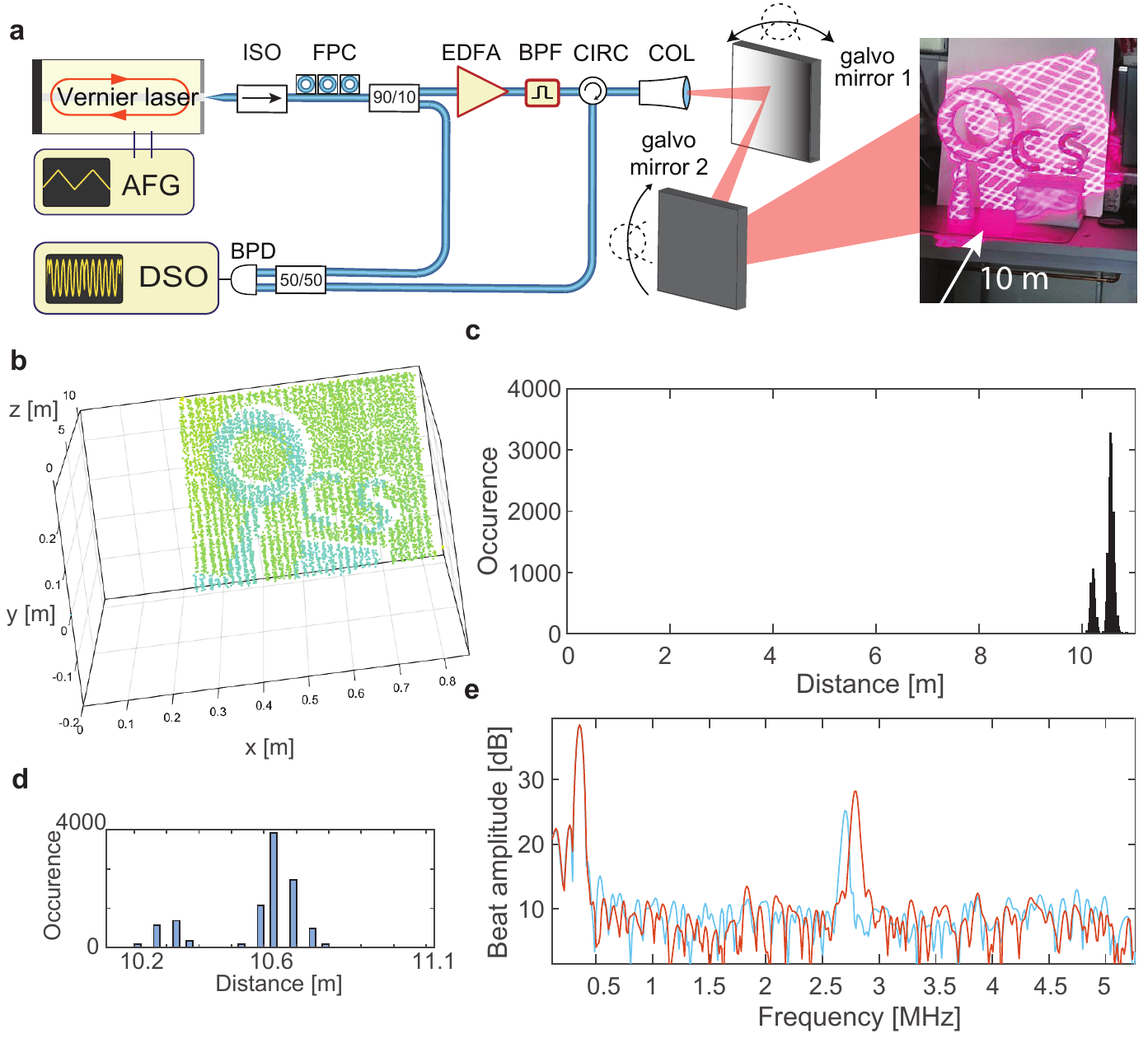}
	\caption{
		\textbf{Optical FMCW LiDAR demonstration.}
		(a)~Schematic of the setup for FMCW LiDAR measurement. ISO - isolator, FPC - fiber polarization controller, EDFA - erbium doped fiber amplifier, CIRC - circulator, COL - collimator, BPD - balanced photodetector, DSO - digital oscilloscope, BPF - optical bandpass filter. 
		A triangular ramp with 11~V peak-to-peak amplitude at 10~kHz rate is applied to both PZT actuators resulting in 1.8~GHz optical frequency excursion of Vernier laser.
		Beam steering is realized using a mechanical galvo scanner with two mirrors at 2~Hz and 63~Hz rates.
		Photos of the target - O,S,C letters 10~m away from the collimator.
		(b)~Point cloud of the target, point colors are based on distance. O,S,C letters in blue, wall in green.
		(c)~Histogram of distance distribution in the point cloud.
		(d)~Histogram of distance distribution near target distances.
		(e)~Processed beatnotes of the local oscillator with the reflections from the collimator at 0.3~MHz and from the target at 2.7~MHz. Red and blue traces correspond to two different time frames. 
	}
	\label{Fig:Fig5}
\end{figure*}

We perform an optical coherent ranging experiment in the lab to demonstrate the potential application of the Vernier laser with tunable \SiN external cavity.
The FMCW LiDAR approach consists of linear frequency modulation of the laser source and delayed homodyne detection with the optical signal reflected from the target.
Figure~\ref{Fig:Fig5}(a) shows the experimental setup of FMCW LiDAR measurement.
The signals driving the PZT actuators are controlled by an AFG.
The laser output after the optical isolator is split to the local oscillator and signal arms with a 95/5 fused fiber splitter.
The signal arm is amplified with an EDFA (Calmar AMP-ST15) between 6~mW and 17 mW, and the ASE noise is suppressed with an optical bandpass filter (Dicon).
Optical beam steering is realized using a mechanical galvo scanner with two mirrors at 2~Hz and 63~Hz.
For the ranging target scene, we use a polystyrene toroid and place additional targets in the form of carton letters "C" and "S" in front of a wall 10~m away from the laser collimator (Fig.\ref{Fig:Fig5}(a)).
We applied a frequency chirp with the predistortion compensation at 10~kHz rate to both PZT actuators to obtain optical frequency excursion of the laser $B = 1.8$~GHz, which corresponds to resolution in distance measurement $c/2B=8.5$~cm, where $c$ is the speed of light.
We record the beat signal of the light reflected from the target and the local oscillator on a balanced PD.
To construct the point cloud from the recorded oscillogram, we first employed a short-time Fourier transform with a window size equal to half of the chirping period and 200\% zero padding.
Fig. ~\ref{Fig:Fig5}(e) depicts the beat note spectra of two different time frames with the reflections from the wall, the letters and the collimator, and their respective signal-to-noise values.
The peaks at 2.7-2.8~MHz offsets correspond to the target scene, while the strong return at 0.3~MHz in Fig.~\ref{Fig:Fig4}(e) is due to the reflection from the collimator.
We find a peak with maximal spectral amplitude in the time-frequency plot for each timeslice, neglecting the collimator reflection and reflections with SNR less than 10 dB.
The frequency of the peak provides the radial coordinate for each timeslice.
Fig.~{\ref{Fig:Fig5}(c,d)} provides a histogram of distance distribution for the target point cloud.
The point cluster at 10.2~m distance corresponds to letters and at 10.6~m to the wall. 
Polar and azimuthal coordinates are retrieved from the galvo scanner mirrors' driving signals, which were recorded on the same digital storage oscilloscope.
Fig.~\ref{Fig:Fig5}(b) depicts the point cloud of the scene with distance-based coloring where the toroid and the letters are depicted in blue and the background wall is depicted in green.

\section{Discussion}
In summary, we have demonstrated a Vernier-filter-based laser using a \SiN PIC with a monolithically integrated PZT actuator and microheater which allows for fast chirp and wavelength switching with <10 ns switching times. 
The combination of cost-effective RSOA and \SiN PIC, along with demonstrated tuning RMS nonlinearity of 0.25\% at 100 kHz chirp rate, make the laser source also well suited for medium to long-range FMCW LiDAR.
Using the same ECL concept but with AlN actuator instead of PZT would allow bringing RMS nonlinearity down to 0.1\% without the need for additional linearization.
Further design improvements might include an intracavity fast phase shifter to increase the laser frequency excursion above 3 GHz, limited in such case by the saturation voltage of the PZT actuator.
The demonstrated approach is based on foundry-ready processes for large-volume photonic integrated circuits and MEMS fabrication, paving the way to the mass production of coherent photonic sensing systems for industrial and consumer applications.

\begin{footnotesize}
	
\noindent \textbf{Funding.} This publication was supported by Contract W911NF2120248 (NINJA LASER) from the Defense Advanced Research Projects Agency (DARPA), Microsystems Technology Office (MTO), as well as the Swiss National Science Foundation (SNSF) through grant number 211728 (BRIDGE). 
A.V. is supported by the EU H2020 research and innovation programme under the Marie Sklodowska-Curie grant agreement No 101033663 (RaMSoM).
J.R. acknowledges funding from the SNSF via an Ambizione Fellowship (No. 201923). 
A.S. acknowledges support from the European Space Technology Centre with ESA Contract No. 4000135357/21/NL/GLC/my.

\noindent \textbf{Acknowledgments.} V.Sn., J.R. and A.V. simulated and designed the devices.
R.N.W. fabricated the device. 
H.T., A.A. designed and fabricated integrated actuators and microheaters.
A.B. and A.V. performed hybrid packaging.
G.L., A.B., V.Sn., J.R., A.V., D.V. characterized the devices. 
J.R., V.Sh. and A.B. implemented linearization of laser tuning.
G.L., A.B., V.Sn, A.V. and J.R. analysed the data, prepared the figures and wrote the manuscript with input from all authors. 
T.J.K., S.B. and A.V. supervised the project.
The chip samples were fabricated in the EPFL center of MicroNanoTechnology (CMi), and in the Birck Nanotechnology Center at Purdue University. PZT layer was fabricated at Radiant Technologies Inc.
We thank Alberto Beccari for helping with the fabrication of glass submounts.

\noindent \textbf{Disclosures.} T.J.K. is a co-founder and shareholder of LiGenTec SA, a foundry commercializing Si$_3$N$_4$ photonic integrated circuits. T.J.K., A.V. are co-founders and shareholders of DEEPLIGHT and A.B. is a shareholder of DEEPLIGHT SA, a start-up company commercializing Si$_3$N$_4$ photonic integrated circuits based frequency agile low noise lasers.

\noindent \textbf{Data Availability.} The code and data used to produce the plots within this work will be released on the repository \texttt{Zenodo} upon publication of the work.

\end{footnotesize}


%

\bibliography{citations}

\end{document}